\documentclass[twocolumn]{aa} % for a preprint version
\usepackage{graphicx,epsfig}
\usepackage{txfonts}
\newcommand{\hi}{\ion{H}{i}}
\newcommand{\oii}{[\ion{O}{ii}]}
\newcommand{\oiii}{[\ion{O}{iii}]}
\newcommand{\kms}{km\,s$^{-1}$}

\newcommand{\dgr}{$^{\circ}$}

\begin{document}

\title{Atomic hydrogen in the one-sided ``compact double'' radio~galaxy \object{\object{2050+364}}}
%\titlerunning{Atomic hydrogen in the one-sided ``compact double'' radio~galaxy \object{\object{2050+364}}}
\titlerunning{\hi \,in the one-sided ``compact double'' radio~galaxy \object{\object{2050+364}}}

\author{R.~C.~Vermeulen \inst{1}
        \and
        A.~Labiano \inst{2,3}
        \and
        P.~D.~Barthel \inst{2}
        \and
        S.~A.~Baum \inst{4,3}
        \and
        W.~H.~de~Vries \inst{5}
        \and
        C.~P.~O'Dea \inst{6,3}
}
\offprints{R.C.~Vermeulen,\hfill\break {\tt rvermeulen@astron.nl}}

\institute{
             Netherl.\ Foundation for Research in Astronomy (ASTRON),
             P.O. Box 2, 7990 AA Dwingeloo, The Netherlands %\\
%             \email{rvermeulen@astron.nl}
         \and %2
             Kapteyn Astronomical Institute, P.O. Box 800,
             9700 AV Groningen, The Netherlands %\\
%             \email{labiano@astro.rug.nl; pdb@astro.rug.nl}
         \and %3
             Space Telescope Science Institute, Baltimore, MD 21218, U.S.A. %\\
%             \email{labiano@stsci.edu; sbaum@stsci.edu; odea@stsci.edu}
         \and %4
             Center for Imaging Science, Rochester Inst.\ of Techn.,
             54 Lomb Memorial Drive, Rochester, NY, 14623, U.S.A. %\\
%             \email{baum@cis.rit.edu}
         \and %5
             Lawrence Livermore National Laboratories, U.S.A. %\\
%             \email{wdevries@igpp.ucllnl.org}
         \and %6
             Department of Physics, Rochester Institute of Technology,
             Rochester, NY, 14623, U.S.A. %\\
%             \email{odea@cis.rit.edu}
}

\date{Received / Accepted}
\authorrunning{Vermeulen et al.}

\abstract{European VLBI Network spectral imaging of the ``compact
double'' radio source \object{2050+364} in the UHF band at 1049~MHz
has resolved the \hi\ absorbing region, and has shown a faint
continuum component to the North (N), in addition to the well-known
East-West double (E, W).

Re-examination of VLBI continuum images at multiple frequencies
suggests that \object{2050+364} may well be a one-sided core-jet
source, which appears as a double over a limited frequency
range. One of the dominant features, W, would then be the innermost
visible portion of the jet, and could be at or adjacent to the
canonical radio core. The other, E, is probably related to shocks at a
sudden bend of the jet, towards extended steep-spectrum region N.

A remarkably deep and narrow \hi\ absorption line component extends
over the entire projected extent of \object{2050+364}.  It
coincides in velocity with the \oiii\ optical doublet lines to within
10~\kms.  This \hi\ absorption could arise in the atomic cores of NLR
clouds, and the motion in the NLR is then remarkably coherent both
along the line-of-sight and across a projected distance of $>300$~pc on
the plane of the sky.

Broader, shallower \hi\ absorption at lower velocities covers only the
%probable 
plausible core area W\null. This absorption could be due to gas
which is either being entrained by the inner jet or is flowing out from
the accretion region; it could be related to the BLR.


}

\maketitle

%________________________________________________________________
%

\section{Introduction \label{sec:intro}}

\object{2050+364} (\object{J2052+3635}, \object{DA 529}; J2000
coordinates $20^{\mathrm h}52^{\mathrm m}52.0549^{\mathrm s}$
$+36^{\circ}35'35.300''$, Beasley et al.~\cite{beasley02}), is one the
original members of the group of radio sources described as ``Compact
Doubles'' (CDs) by Phillips \&\ Mutel (\cite{philmut81},
\cite{philmut82}), and then studied in more detail by Mutel, Hodges,
\&\ Phillips (\cite{mutel85}). Phillips \&\ Mutel (\cite{philmut81})
suggested that, in contrast to most of the radio-loud AGN imaged with
VLBI at GHz frequencies, CDs are ``mini-lobes'', on (sub)galactic
scales, at the ends of bi-directional outflows from an unseen central
(``core'') component. The spectrum of the two components in compact
doubles is typically not flat but peaked at a frequency around a GHz;
such sources are termed Gigahertz Peaked Spectrum (GPS) sources (see
the review by O'Dea \cite{odea98}). The spectral shape can often be
reasonably well understood from synchrotron self-absorption in
$\sim1$~GHz radio emitting regions of $\sim100$~mas cross-section, but
there are indications that in some sources, the low-frequency spectral
shape is also determined (in part ?) by free-free absorption (e.g.,
Bicknell, Dopita, \&\ O'Dea \cite{bicknell97}; Kameno et
al.~\cite{kameno03}, hereafter K03; Risaliti, Woltjer, \& Salvati
\cite{risaliti03}).

The CD and GPS source classes are related to the group of Compact
Symmetric Objects (CSOs), first so termed by Conway et al.\
(\cite{conway94}) and Wilkinson et al.\ (\cite{wilkinson94}). In CSOs,
a central compact core component is visible in between two lobe-like
components, albeit usually only faintly and often only at high
observing frequencies. Thus, in CSOs the two-sidedness of the radio
emission is proven beyond reasonable doubt. Models have been considered
in which the lobes are kept to subgalactic dimensions by a dense
confining medium (e.g., De Young \cite{deyoung93}; Carvalho
\cite{carvalho94}, \cite{carvalho98}). However, in a number of CSOs
(e.g.\ Owsianik \&\ Conway \cite{owsianik98}; Polatidis \&\ Conway
\cite{polatidis03}; Gugliucci et al.\ \cite{gugliucci05}) lobe advance
velocities of a few tenths $c$ have now been observed, and this
provides powerful evidence that CSOs structures signify young radio
sources, some with kinematic ages as low as a few hundred years.

GPS radio sources at moderate redshifts ($z<1$) are typically
identified with passively evolving elliptical galaxies (e.g., Snellen
et al.\ \cite{snellen98}). The lifecycle of these sources is a subject
of current interest, and is not well understood: at least some of them
probably evolve to become fully fledged double-lobed FR\,I or perhaps
even FR\,II radio sources, which typically have linear sizes of 100~kpc
or even up to 1~Mpc, although, based on number counts, they should
decrease in luminosity as they grow (e.g., Fanti et al.\
\cite{fanti95}; Readhead et al.\ \cite{readhead96}; O'Dea \&\ Baum
\cite{odeabaum97}; Alexander \cite{alexander00}; Snellen et al.\
\cite{snellen00}). On the other hand, perhaps some compact sources may
have multiple very short-lived active episodes (e.g., Gugliucci et al.\
\cite{gugliucci05}). Compact double radio sources are also particularly
promising objects to study in the context of AGN fuelling and radio
source evolution models, and the interaction between the interstellar
medium and jets, because they lie entirely within the inner galaxy, on
kpc or even sub-kpc scales.

One attractive possibility is to observe the 21cm hyperfine line of
atomic hydrogen in absorption against bright radio structure, to study
the kinematics and density distribution of the atomic gas. VLBI offers
spatial resolution on the scales of the NLR and sometimes even the BLR,
not accessible by any other means in galaxies at any appreciable
redshift. However, for most CPS/CSO/CSS radio sources the frequency of
this \hi\ line is redshifted outside the traditional observing
band. But profiting from the advent of UHF receivers on
interferometers, first at the Westerbork Synthesis Radio Telescope
(WSRT), and then at many telescopes in the European VLBI Network (EVN),
it has become possible to study the presence and distribution of
associated \hi\ in absorption against many compact radio sources.

A WSRT survey of a sample of compact radio sources was published by
Vermeulen et al.\ (\cite{verm03}, hereafter V03), and further analysed
by Pihlstr\"om, Conway, \& Vermeulen (\cite{pihl03}). About one third
of the objects surveyed had detectable \hi\ absorption, and the peak
depth was found to be anti-correlated with linear size. A 16~\%\ peak
depth HI absorption line at $z=0.3547$\footnote{H$_0=70$
km$\,$s$^{-1}\,$Mpc$^{-1}$, $\Omega_{\mathrm m}=0.27$, and
$\Omega_{\Lambda}=0.73$ are adopted throughout this paper, so that an
angular size of 1~mas corresponds to 5.0~pc at $z=0.3547$.} was
discovered towards \object{2050+364}. This is uncommonly deep; for
sources of similar overall linear extent, absorption depths of
0.1--1~\%\ (or upper limits at that level) were found to be rather more
typical by V03. \object{2050+364} was therefore included in one of the
first sessions with the new UHF receivers on the EVN, in order to
resolve the HI absorption region spatially. We describe the data
analysis procedure in Sec.~\ref{sec:obs}, and present the results in
Sec.~\ref{sec:res}.

\object{2050+364} has received comparatively little attention in
its own right during the last two decades. Perhaps this was because
of its low Galactic latitude, $b=-5$\dgr, in the Cygnus superbubble
region ($l=79$\dgr), which is a handicap for obtaining a secure optical
identification because of crowding, and hampers low frequency radio
studies of the object because its image is broadened by interstellar
scattering (Dennison et al.~\cite{dennison84}, Mutel \&\ Hodges
\cite{mutel86}). Biretta, Schneider, \&\ Gunn (\cite{biretta85}) and
O'Dea, Baum, \& Morris (\cite{odea90}) have both found a good match
between the radio position and an $m_{\mathrm r}=21.1$, $(r-i)=0.1$
galaxy. A spectrum was published by de Vries et al.\ (\cite{devries00})
showing prominent H$\beta$ and \oiii\ emission lines. The latter have a
redshift $z=0.3547$; we will critically review the redshift of
\object{2050+364} in Sec.~\ref{subsec:dis-kin} and the optical
identification in Sec.~\ref{subsec:dis-loc}. Meanwhile, the low
galactic latitude of \object{2050+364} has been turned into an
advantage by using it as a background probe for interstellar
scattering (e.g., Fey \&\ Mutel \cite{fey93}). This has
culminated in a multi-frequency VLBI imaging study by Lazio \&\ Fey
(\cite{lazfey01}, hereafter LF). Another multi-frequency VLBI
dataset is available from the investigation into possible free-free
absorption in a sample of GPS sources by K03. Their results are
described in Sec.~\ref{sec:multi}, and used in our discussion on the
nature of \object{2050+364} in Sec.~\ref{sec:dis}.

\section{Observations and data processing}\label{sec:obs}

\subsection{Observations}\label{subsec:obs}

\begin{figure}[thb]
\centering
\centerline{
\includegraphics[angle=0,width=8.5cm]{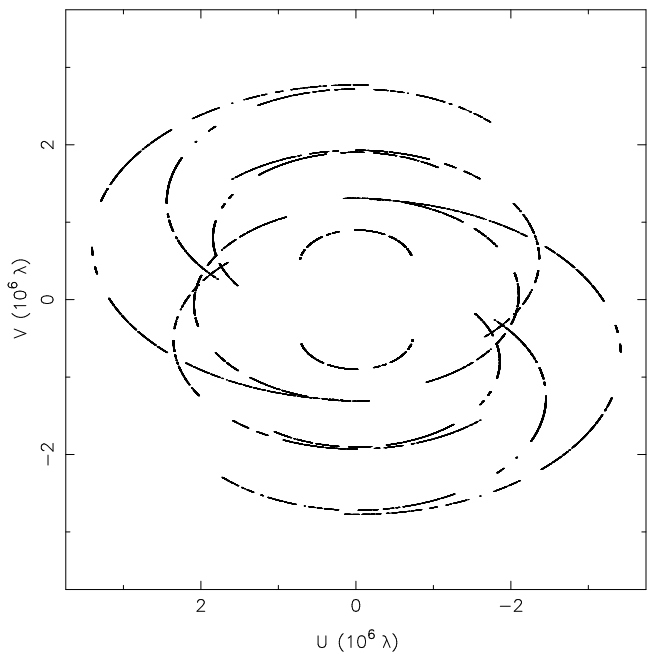}
}
\caption{The (u,v)-coverage obtained on \object{2050+364} with the EVN
  telescopes Effelsberg, Jodrell Bank, Onsala, and WSRT\null. Crossing
  points provide important self-calibration constraints.}
\label{fig:uvcov}
\end{figure}

On 1999 September 08--09, in one of the earliest sessions at UHF
frequecies with the European VLBI Network (EVN), 14 hours were spent
observing \object{2050+364}, and the calibrators \object{DA\,406},
\object{3C\,454.3}, and \object{3C\,84}.  Four telescopes produced
usable data: dual circular polarisations were available at Effelsberg
and the WSRT, while Jodrell Bank had only LCP, and Onsala recorded dual
linear polarisations. The (u,v)-coverage on \object{2050+364} is
shown in Fig.~\ref{fig:uvcov}. The observing band, 1046.6~MHz to
1050.6~MHz, was centred on the absorption line discovered with the WSRT
by V03. A more extensive description of early EVN UHF observing
characteristics and data analysis procedures will be given in Vermeulen
et al.\ (in preparation).

\subsection{Initial calibration and data averaging}\label{subsec:cal}

The initial data processing (fringe fitting, spectral passband
calibration, and a priori complex gain calibration) took place in the
NRAO Astronomical Image Processing Software (\verb|AIPS|), using the
calibrator sources.  After time-interpolated transfer of the solutions,
and residual fringe-fitting on \object{2050+364}, the data
were averaged into 256 independent spectral channels (each 4.5 \kms\
wide), and 60 sec time samples. This increased the sensitivity per
visibility. The target source, \object{2050+364}, was detected on all
baselines.

All available polarisation products were also averaged together,
including the cross-correlations between the linear polarisations from
Onsala and the circular polarisations from the other telescopes. We
believe the sensitivity gained to total intensity is more important
than any possible resultant limitations on absolute flux calibration
accuracy or on (image or spectral) dynamic range; we think these are
more affected by the sparseness of the array and the lack of complete
system temperature data and gain curves.

\begin{figure}[thb]
\centering
\centerline{
\includegraphics[angle=0,width=8.5cm]{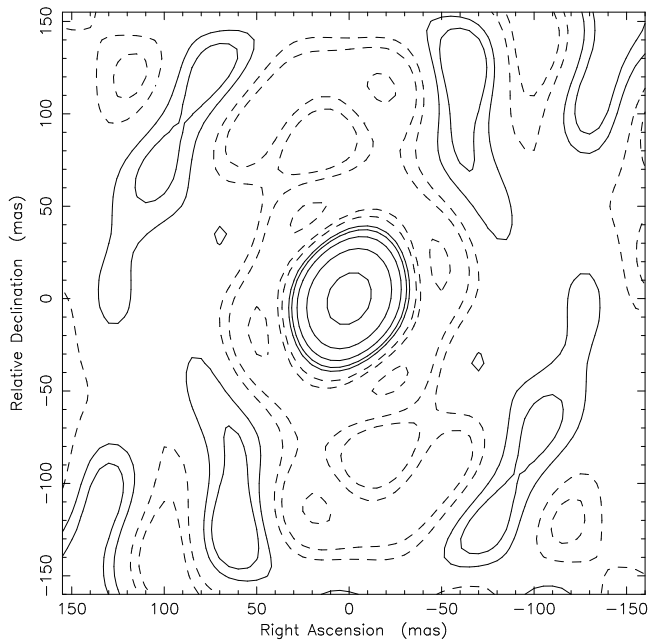}
}
\caption{The dirty beam obtained with the (u,v)-coverage shown in
  Fig.~\ref{fig:uvcov}; contour levels are drawn at $-0.2,\ -0.1,\
  -0.05,\ 0.05,\ 0.1,\ 0.2,\ 0.4,\ {\rm and}\ 0.8$. This beam is benign for
  deconvolution of the continuum structures in Fig.~\ref{fig:image}}
\label{fig:dbeam}
\end{figure}

Much care was taken to obtain consistent visibility calibration, for
the rather sparse array, and in the occasional presence of external
radio interference (although its impact is mitigated since it is
typically not coherent over VLBI baselines). Relative antenna gains
were derived from requiring consistent visibility amplitudes on the
calibrator sources, while the overall flux density scale was set such
that in the final self-calibrated image (see below) the total flux
density is equal to 3.34~Jy at 1049~MHz, a value derived by
interpolation from total flux densities at a number of other
frequencies (Salgado et al.\ \cite{salgado99}; White \&\ Becker
\cite{white92}). We estimate the overall uncertainty in the flux scale
of this UHF VLBI dataset to be as much as 20~\%, but the main
astrophysical results are unaffected, since they depend on opacities
rather than on absolute flux densities.

\begin{figure}[thb]
\centering
\centerline{
\includegraphics[angle=0,width=8.5cm]{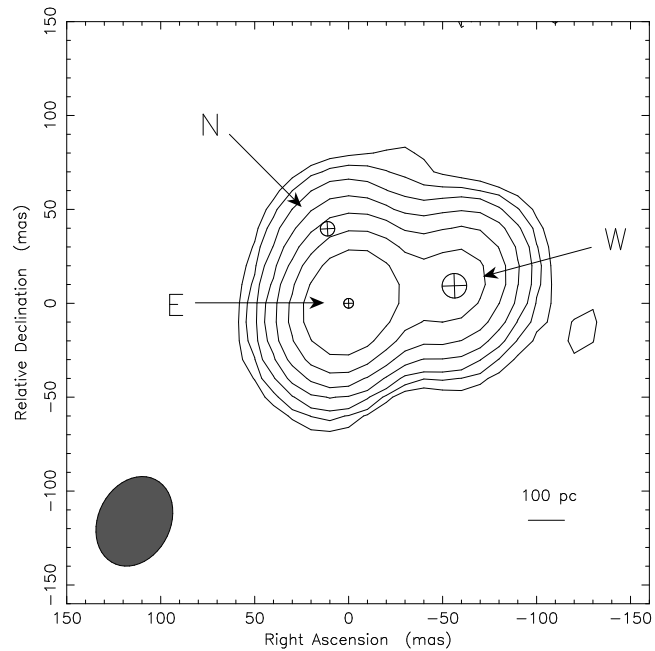}
}
\caption{Continuum VLBI image of \object{2050+364} at 1049~MHz. The
  restoring beam is $50\times38$~mas in position angle $-27$\dgr. The
  peak flux density is 2.42~Jy/beam; contours start at 0.015~Jy/beam
  (3 times the r.m.s.\ fluctuation level away from the continuum
  structure), and increase in factors of 2. The absolute flux density
  scale is uncertain by 20~\%, but this does not affect relative flux
  densities; see Sec.~\ref{subsec:cal}. The three circular Gaussian
  components used for modeling (Table~\ref{tab:model}) are
  overplotted.}
\label{fig:image}
\end{figure}

\subsection{Self-calibration; continuum imaging and
modeling}\label{subsec:cont}

The Caltech \verb|DIFMAP| software package (Shepherd, Pearson \& Taylor
\cite{shepherd}) was used for all further calibration and analysis. A
continuum dataset was formed by averaging all of the line-free channels
on both sides of the \hi\ absorption line, of which the location and
width were taken from the WSRT survey (V03). 
%Despite the sparseness of the VLBI array available, 
The available VLBI array was sparse, but contained some crossing
points (see Fig.~\ref{fig:uvcov}), and careful self-calibration was
feasible. A double source structure was already evident in the first
image made. After initial phase self-calibration, some additional flux
density to the north-east quickly became apparent. At that time, we
were unaware of the images published by LF, which at the lower
frequencies also show emission in that area. Extensive tests of the
reality of this third component were convincing:
%its presence allows for more self-consistent and less time-variable
%self-calibration solutions.
our data demand its presence. There are no major sidelobes of the
dirty beam (shown in Fig.~\ref{fig:dbeam}) to affect its
deconvolution with respect to the brighter parts of the source. We have
performed careful iterative cycles of self-calibration with
imaging/cleaning/modeling, both in trial runs where we kept excluding
the third component from the models, and in runs where we admitted the
presence of this additional continuum feature. The latter both gave
cleaner-looking images and, more significantly, consistently required
less extreme and more time-stable self-calibration coefficients.

\begin{table} \begin{center} 
  \caption{The three circular Gaussian components used for
  modeling. Their continuum flux densities are listed. The absolute
  flux density scale is uncertain by 20~\%, but this does not affect
  relative flux densities; see Sec.~\ref{subsec:cal}. These components
  are overplotted on the image in Fig.~\ref{fig:image}.}
 \begin{tabular}{c c c c c c c}
    \hline
    \hline
     Component  &Flux &Radial      &P.A.  &Diam \\
                &(Jy) &dist (mas)  &(deg) &(mas) \\
    \hline
    W & 0.79 &  0 &   0 & 13.1 \\
    E & 2.44 & 57 &  99 & 5.2  \\
    N & 0.11 & 74 &  66 & 7.7  \\
    \hline
  \end{tabular} 
  \label{tab:model}
  \end{center}
\end{table}

In order to restrict the number of free parameters, the sky model
fitted to the visibility data during iterative self-calibration cycles
consisted of three circular Gaussian components: W(est), E(ast), and
N(orth); their parameters, fitted to the visibility data, are given in
Table~\ref{tab:model}. The final self-calibrated continuum visibility
data were used to produce the cleaned and restored image displayed in
Fig.~\ref{fig:image}; symbols showing the three model components are
overplotted.

\subsection{Spatially resolved spectroscopy}\label{subsec:line}

The sensitivity of the data and the sparseness of the array did not
allow generation of a full spectral image cube of acceptable
signal-to-noise ratio. Instead, spectra of the W, E, and N areas of the
source were determined by re-fitting the flux density of each of the
three Gaussian model components derived for the line-free continuum,
but then separately for each of the 256 spectral channels. The
positions and diameters of the three components were not varied; they
were fixed at the values derived from the continuum.

\begin{figure}[thb]
\centering
\centerline{
\includegraphics[angle=0,width=8.5cm]{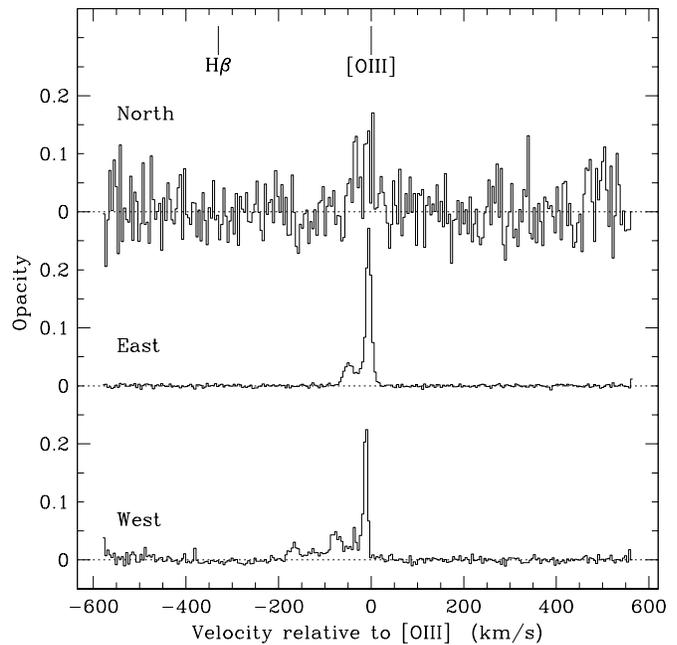}
}
\caption{The \hi\ absorption spectra, expressed in opacity, for the W,
  E, and N radio components of \object{2050+364}. Zero velocity is chosen to be
  the \oiii\ redshift, $z=0.3547$ (see Sec.~\ref{subsec:dis-kin}).}
\label{fig:overview}
\end{figure}

\begin{figure}[thb]
\centering
\centerline{
\includegraphics[angle=-90,width=8.5cm]{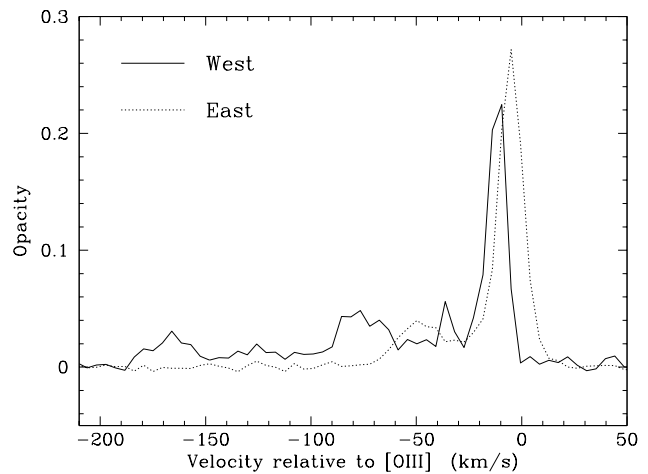}
}
\caption{Enlargement of part of Fig.~\ref{fig:overview}, to allow
  detailed comparison of the \hi\ opacity profiles towards W and E.}
\label{fig:detail}
\end{figure}

\begin{figure}[thb]
\centering
\centerline{
\includegraphics[angle=0,width=8.5cm]{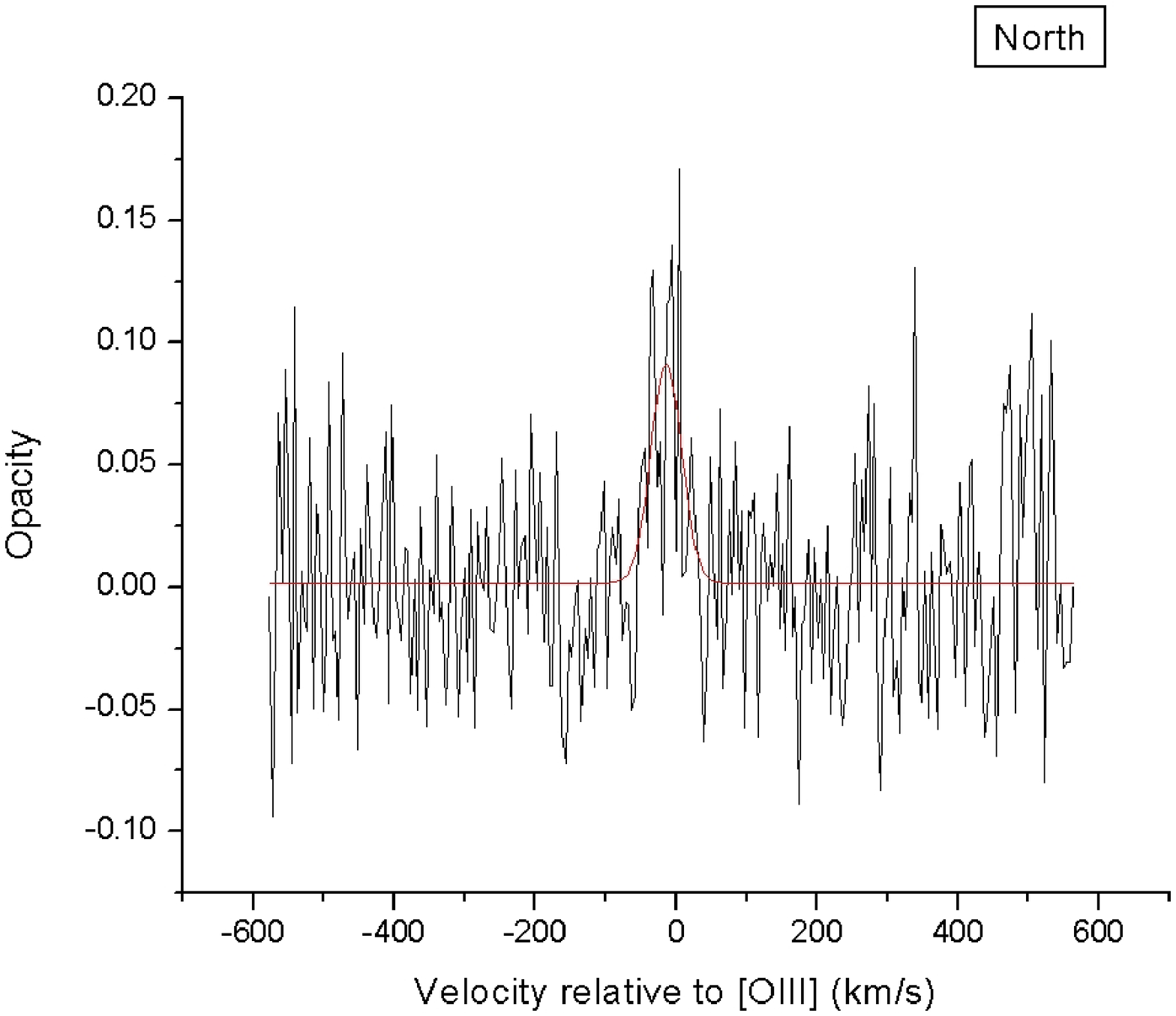}
}
\centerline{
\includegraphics[angle=0,width=8.5cm]{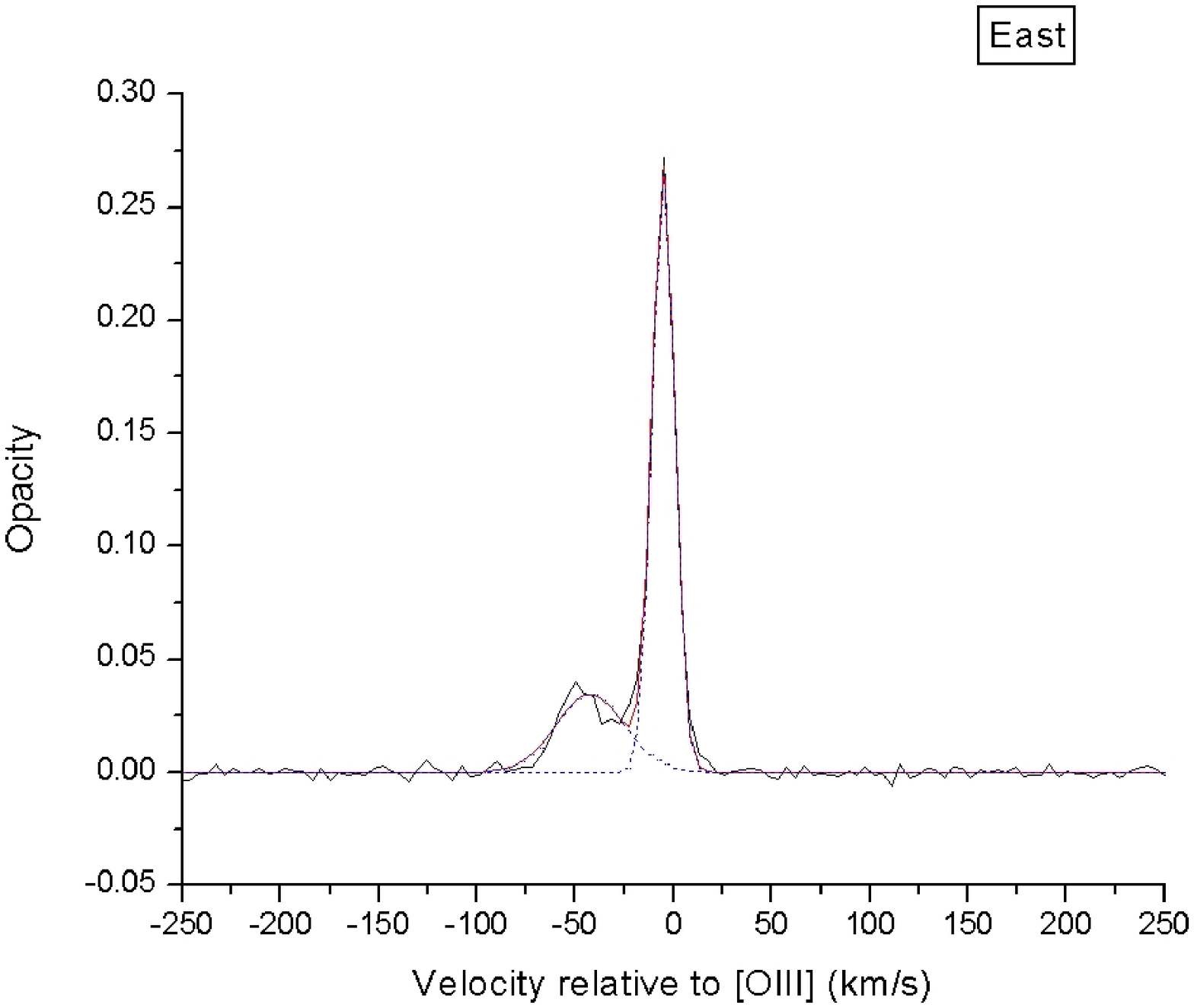}
}
\centerline{
\includegraphics[angle=0,width=8.5cm]{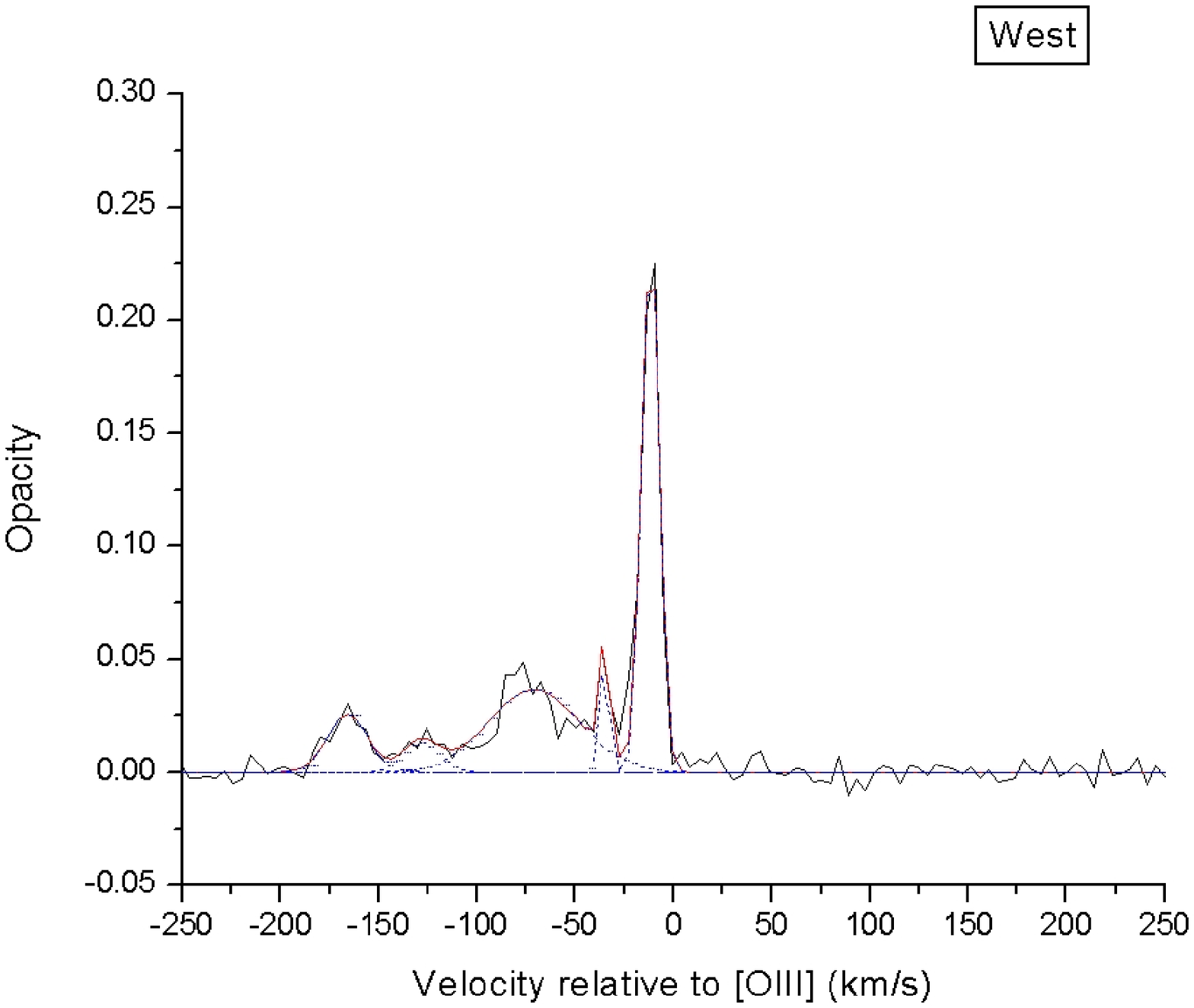}
}
\caption{Fitted Gaussian line profiles overplotted onto the \hi\
absorption spectra of the three regions. The parameters are listed in
Table~\ref{table:linegauss}.}
\label{fig:linegauss}
\end{figure}

The resultant spectra are shown in Figures \ref{fig:overview} and
\ref{fig:detail}, as opacity with respect to the continuum strength of
the appropriate components; this avoids the uncertainties in the
absolute flux density scale, discussed in Sec.~\ref{subsec:cal}. Zero
velocity for Figs.~\ref{fig:overview} and \ref{fig:detail} corresponds
to the redshift of \oiii, $z=0.3547$, published by de Vries et al.\
(\cite{devries00}); this choice will be further discussed in
Sec.~\ref{subsec:dis-kin}. Note that a less precise value of the
optical redshift was used by V03.

\begin{table}
\begin{minipage}{\columnwidth}
\caption{The parameters of the Gaussian line profiles fitted to the
observed \hi\ spectra of the three regions, as shown in
Fig.~\ref{fig:linegauss}. The derived column depths assume uniform
coverage and $T_\mathrm{sp}=100$~K.}
\label{table:linegauss}
\centering
\begin{tabular}{crrcc}
\hline
\hline
Region&Centre&FWHM&Peak opacity& $N_\mathrm{\hi}$\\
  & (\kms) & (\kms) & &10$^{20}$ cm$^{-2}$\\
\hline
W &  $-$11.7 $\pm$ 0.1  & 10.3 $\pm$  0.3 & 0.212 $\pm$ 0.010 & 4.2 $\pm$ 0.2\\
     &  $-$35 0 $\pm$ 0.7  &  5.8 $\pm$  2.1   & 0.037 $\pm$ 0.022 & 0.4 $\pm$ 0.3\\
     &  $-$71.0 $\pm$ 2.3  & 55.0   $\pm$ 6.9   & 0.036 $\pm$ 0.006 & 3.9 $\pm$ 0.8\\
     & $-$128.8 $\pm$ 3.6 & 20.7   $\pm$  9.2   & 0.013 $\pm$ 0.008 & 0.5 $\pm$ 0.4\\
     & $-$166.1 $\pm$ 1.9 & 24.1   $\pm$  4.8   & 0.026 $\pm$ 0.007 & 1.2 $\pm$ 0.4\\
E  & $-$5.1  $\pm$ 0.1  & 13.4 $\pm$  0.2 & 0.263 $\pm$ 0.004 & 6.8 $\pm$ 0.1\\
    &  $-$42 .4  $\pm$ 0.9 & 41.1   $\pm$  2.4   & 0.034 $\pm$ 0.003 & 2.7 $\pm$ 0.3\\
N &  $-$14.2   $\pm$ 5.1 & 51.5 $\pm$12.1   & 0.091 $\pm$ 0.028 & 9.1 $\pm$ 3.6\\
\hline
\end{tabular}
\end{minipage}
%\\ \\
%Summary of the fitted values to absorption features in the different 
%regions of \object{2050+364}. The first column identifies the region 
%and absorption component in that region. Center gives the velocity of 
%the peak of the absorption. The third column gives the fitted peak 
%opacity and the last column the column density for that opacity and 
%FWHM.\\
\end{table}

Gaussian line profiles were fitted to the most obvious features of the
\hi\ spectra at the three locations, using Origin 6.1. The fitted
profiles are overplotted on the data in Fig.~\ref{fig:linegauss}, and
their parameters are listed in Table~\ref{table:linegauss}. These serve
only to obtain a rough quantification of the absorber properties; in
particular, no attempt was made to find an optimally fitting set of
Gaussians to cover the complex observed line spectrum at W.

\section{Results}\label{sec:res}

\subsection{VLBI continuum structure of \object{2050+364} at
1~GHz}\label{subsec:rescon}

The 1049~MHz continuum VLBI image shows the W and E components, which
constitute the well-known ``classical double''. At this frequency the
W:E flux density ratio is close to 1:3; this is constrained better than
the flux densities themselves. The two components are separated by
about 57~mas (285~pc) along a PA of $\sim$100\dgr, which matches the
relative locations found at other frequencies by LF. The component
sizes we find at 1~GHz are only slightly larger than those found at
1.67~GHz by LF; we believe that the accuracy of our fitted component
sizes is limited by the relatively poor resolution of our array
(synthesised beam $50\times38$ mas), although it is interesting that
the angular diameter we find for E, 5.2 mas, is comparable to the
scattering disk diameter determined by LF, 6--8 mas (as discussed
below, this could be somewhat of an over-estimate since LF did not
recognise sub-components which exist at the lowest frequencies.

We have also discovered with our data (before finding it in the
low-frequency images published by LF) an additional emission region to
the north/northeast, which we call N\null. While we are confident
about the existence and approximate location of N, about 74~mas
(370~pc) from W in PA $\sim66$\dgr, its flux density ($\sim$0.1~Jy)
is subject to considerable uncertainty, both in view of the overall
flux scale uncertainty and because, with the sparse imaging array, some
extended emission may have been missed. The relatively small array of
EVN antennas used did not allow the presence of substructure within any
of the three components to be investigated.

\subsection{\hi\ absorption distribution}\label{subsec:resspec}

The VLBI data reveal that the two most prominent \hi\ absorption
features (16~\%\ and 4~\%\ integrated opacity) found towards
\object{2050+364} in the WSRT survey (V03) do not cover the sub-kpc
scale radio source uniformly. The integrated (WSRT) profile is
dominated by the absorption towards E, because that component is much
brighter than the other radio source components.

The distinct, deep absorption feature at the high velocity end of the
absorption spectrum shows a $\sim$7~\kms\ centroid velocity offset
between W and E; see also Fig.~\ref{fig:detail}. This observed
offset is highly significant, given the velocity resolution of
4.5~\kms, and the high signal-to-noise ratio of these absorption
features. The line is also narrower at W than at E\null.  At N, a
corresponding absorption feature is also visible in the VLBI data (see
Figs.~\ref{fig:detail} and \ref{fig:linegauss}), but, due to the low
background continuum flux density, the centroid velocity and the FWHM
are too uncertain for a useful comparison. The peak opacity at N
appears to be lower than at either W or E\null.

The VLBI data show that the second absorption feature tabulated by V03
for the integrated WSRT spectrum (where it covers the observed
velocity range $-30$~\kms\ to $-60$~\kms\ with respect to the \oiii\
redshift) has a peak opacity of $\tau=0.035$ at both W and E\null. But
the detailed profile differs substantially between W and E; it is
particularly irregular at W\null. The integrated profile shown in V03
is dominated by E, because it is the brightest feature. While the
absorption at N was modeled with just a single Gaussian line, it is
wider than the main component at E and W (see
Table~\ref{table:linegauss}, and may therefore also encompass some
secondary absorption features.

The VLBI data also show that there are further absorption features,
extending to an observed velocity of nearly $-200$ \kms, at
comparatively low opacities (a few percent, see
Fig.~\ref{fig:linegauss}). These are all detected towards W only, and
not towards E; towards N the signal-to-noise ratio is too poor to
establish or delimit the presence of similarly shallow features. Since
E dominates the total flux density, these features are not readily
apparent in the integrated spectrum of V03.

\section{Multi-frequency Radio Continuum Structure}\label{sec:multi}

The compact structure of \object{2050+364} was imaged with VLBI at
multiple frequencies by both LF and K03.
%That project  was aimed at studying interstellar scattering,
Their projects were aimed at studying interstellar scattering, or
free-free absorption statistics, respectively, and the structure of
%the source 
\object{2050+364} itself was not analysed in detail. However, we
give our own description of the images
%of LF
here, because we will then show in Sec.~\ref{sec:dis} that
together with our own results, a surprising new interpretation emerges:
\object{2050+364} could well be a one-sided core-jet source !

\begin{figure}[thb]
\centering
\centerline{
\includegraphics[angle=-90,width=8.5cm]{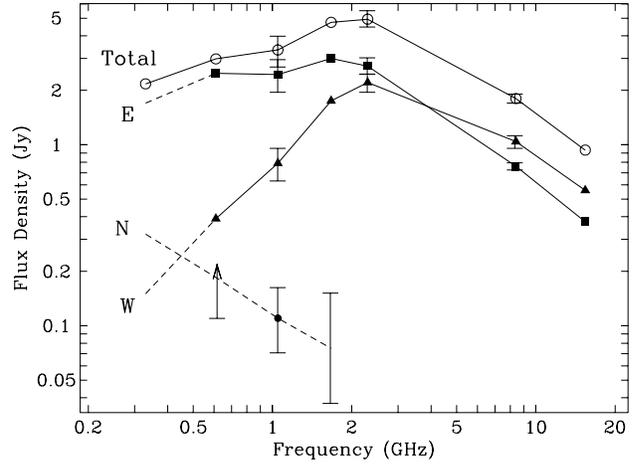}
}
\caption{Continuum component flux densities at multiple
  frequencies, based on data at 0.33, 0.61, 1.67, 2.3, and 8.4~GHz from
  LF, at 2.3, 8.4, and 15.4~GHz from K03, and at 1.049~GHz from our own
  observations. See Sec.~\ref{sec:multi} for details.}
\label{fig:lf}
\end{figure}

To support the discussion, we show in Fig.~\ref{fig:lf} the flux
densities at multiple frequencies as tabulated by LF and K03, and
the flux densities at 1~GHz from our own data
(Table~\ref{tab:model}). We have summed over all sub-components which
were given the same label (W or E, respectively) by LF, except, as
discussed below, at 0.61~GHz, where we recognise one of the components
as N, and at 0.33~GHz, where we think the single flux density listed by
LF should in fact be ascribed to a combination of all components.
LF and K03 both have data at 2.3~GHz and 8.4~GHz. The morphologies in
their images appear to match well, and while their tabulated flux
densities differ at the 20~\%\ level, the uncertainties do not obscure
the trends on which we wish to focus; we simply use the average of the
two available measurements in Fig.~\ref{fig:lf}, with a bar to show the
range between them.

The northeastern region, which we have called N, is clearly
distinguishable in the 1.67~GHz image of LF, where it consists of
rather extended emission at a distance of roughly 30--55~mas from
region E, to the north and northeast, in position angles roughly
between 0\dgr\ and 30\dgr\ as seen from E; the position angle of region
N as seen from W is about 65\dgr. Note that this corresponds well with
the position of N in our data (see Table~\ref{tab:model}; with respect
to E, the distance of N at 1049~MHz is 41~mas, in a position angle of
16\dgr). There are, understandably, no model components tabulated by LF
for this extended emission. From the contour levels and extent of the
emission, our very rough estimate of the total flux density of N is
75~mJy, which we have indicated with a factor of 2 margin towards both
lower and higher values in Fig.~\ref{fig:lf}.

Region N is also unmistakably present in the 0.61~GHz image of LF.
Even though, for their own analysis, LF have included it as part of E,
the third 0.61~GHz model component tabulated by LF (with flux density
0.11~Jy) clearly forms part of N, given its position relative to
E\null. The model component might only represent part the total flux
density from that area, and it is thus indicated as a lower limit in
Fig.~\ref{fig:lf}.

At 0.33~GHz, LF have modeled the source as a single Gaussian, which
they ascribed fully to E\null. However, in their image we find good
evidence for the presence of all three components, W, E, and
N\null. The image is clearly extended and elongated, and the western
side is the fainter part. We believe this is W; all other data
% of LF
establish that W has a more inverted broad-band spectrum and is fainter
than E at all frequencies 
%except 
below 8.4~GHz (see
Fig.~\ref{fig:lf}). Unfortunately, E and N are difficult to disentangle
in the 0.33~GHz image, both because of the lower VLBI resolution, and
because of scatter-broadening, although LF may have overestimated the
amount of interstellar scattering at 0.33~GHz when they took the entire
visible structure to be due to E\null. We find that the overall
position angle of the contour image is 75\dgr--85\dgr. This suggests
that the location of the centroid of the bright, eastern part of the
source is influenced by significant contributions from both E (in
PA$\sim$100\dgr\ from W) and N (in PA$\sim$65\dgr\ from W). Thus, we
show in Fig.~\ref{fig:lf} a plausible but by no means unique
decomposition of the total flux density of the single model component
fitted by LF; to indicate our uncertainty, no marker symbols are drawn.

Components W and E, we believe, are thus separately visible
%in the images of LF
at all frequencies, while component N is visible in the images up
to 1.67~GHz. Above 1~GHz, where the linear resolution is sufficient
and the impact of interstellar scattering negligible, ``component'' E
shows considerable substructure. At the highest observing
%frequency of LF, 8.42~GHz, 
frequencies, 8.4~GHz and 15.4~GHz,
``component'' W is seen to 
%be a double 
have compact substructure as well. Where measurable, the
intrinsic angular sizes of W or its sub-components are considerably
smaller than those of E or its sub-components. Also, from
Fig.~\ref{fig:lf}, W clearly has the most inverted low-frequency
spectrum, and very likely also the highest peak frequency and the
flattest high-frequency spectrum.

N has quite the opposite properties: it is the most extended and does
not appear to have much compact substructure; its broad-band radio
spectrum is probably steep over the full frequency range (although with
large uncertainties from the data of LF). Thus, W, E, and N,
successively are in a sequence of decreasing compactness, steepening
high frequency spectral index, and decreasing peak frequency.

\section{Discussion \label{sec:dis}}

\subsection{\object{2050+364}: a one-sided core-jet structure ?\label{subsec:dis-str}}

% This gradual shift of % properties 
The three regions W, E, and N show a progressive change of
compactness and of spectral shape which is not typical for compact
double-lobed radio sources, but rather for core-jet radio sources, where
relativistic beaming and opacity effects lead to a one-sided observed
structure. Furthermore, the relative alignment of the three
regions is also compatible with a one-sided core-jet source. The 
very compact sub-components of W seen at
%8.42~GHz 
the highest frequencies are aligned roughly, but not exactly,
along the line towards E\null. 
% Furthermore
The relative disposition of the substructures in E at various
frequencies suggests that this may well be radio emission from
shocks at the location of a sudden bend in a single jet, which
%originates at 
is first visible at W, and after bending at E continues
towards the more extended emission region N\null. The jet could be at a
fairly small angle to the line of sight, such that the deprojected
angle of the bend at E need not be large; this is seen in many other
core-jet sources. Due to its substructure, E has a somewhat unusual
convex broad-band radio spectrum, peaked at frequencies of a few GHz.
We find that \object{2050+364} is set apart from many other
GPS/CSS sources by the fact that at all GHz frequencies its total flux
density is dominated by two emission regions, W and E, which are rather
compact and widely separated, i.e., each one spans only a small
fraction of the distance between them. Genuine double-lobed GPS/CSS
sources can show compact hot spots at the two ends, but then in
addition they typically have clearly visible, often dominant, extended
emission regions (the lobes). Region N in \object{2050+364}, on the
other hand, is rather faint, even at the lowest frequencies, and as
such would be rather atypical if it were a lobe beyond its putative
hot-spot E\null. A convincing counter-lobe is not seen. Classical
core-jet sources, on the other hand, do often show a faint extended
emission region beyond a shock at a bend in the jet.

We think that it is, furthermore, plausible that the
%two 
sub-components of W
%are 
may be at or adjacent to the traditional radio core. Many core-jet
radio sources show a few bright knots in the innermost jet. The
overall apparent separation velocity between W and E at 1.67~GHz has
been delimited by LF to be no faster than $c$. However, this limit is
still several times faster than what seems to be typical in CSOs (e.g.,
Gugliucci et al.\ \cite{gugliucci05}), and it also does not rule out
that some sub-components might even show superluminal motions, as is
sometimes seen in other sources, where stationary radio components are
thought to mark regions of either confinement or bending in a jet,
while other radio features may be moving with a flow or due to
traveling disturbances. It is possible that the sub-components of W are
still part of the inner jet, with the true core being somewhat further
to the west, perhaps at a more compact radio component, as yet unseen
due to a spectrum which is significantly inverted as a result of
synchrotron self-absorption, free-free absorption, or a combination of
both, as is probably the case in several other compact radio sources
(e.g., Kameno et al.\ \cite{kameno03}). It would be interesting to
attempt to monitor the positions of some of the well-defined
sub-components in the W and E regions at high frequency.

In most well-known two-sided GPS/CSS sources, at the relatively
high frequency of 15 GHz synchrotron self-absorption and/or free-free
absorption do not fully hide the central core component (to a
reasonable dynamic range of a hundred, say). However, even higher
frequency observations would be useful before definitely ruling out
the existence of a more inverted-spectrum core component between W and
E, in which case the radio source might be two-sided after all.
Likewise, even if the currently imaged structure is one-sided, with the
core at or near W, it is possible that further high-dynamic-range
imaging could reveal emission from a second jet or lobe on the opposite
side. This situation has occurred, for example, in 2352+495, a
well-known CSO (see Conway et al.~\cite{conway92}).

The original selection of \object{2050+364} as a Compact Double
(suggesting two-sided radio emission) seems to have been the result of
a similarity between W and E which is partly coincidental; we now know
that their properties diverge when studied over a wider range of
frequencies, and that a section further out in the jet, N, has a
steeper, more typical broad-band spectrum.  Several other examples are
now known in which the similarity between the components of an apparent
compact double breaks down when the source is studied with high enough
resolution and dynamic range and at high enough frequency. Amongst the
sources now considered to be one-sided is \object{CTD93} (see Shaffer,
Kellermann, \&\ Cornwell \cite{shaffer99}); like \object{2050+364}, it
was a member of the original group of CDs. The identification of a
centrally located core component is a crucial part of confirming
whether a CD or GPS source is indeed two-sided, and therefore a CSO,
rather than a one-sided core-jet source in which two emission regions
happen to have comparable flux densities and sizes, and therefore
roughly similar spectra (Conway et al.~\cite{conway94}, Wilkinson et
al.~\cite{wilkinson94}).

\subsection{The optical redshifts and \hi: infall or outflow ?
\label{subsec:dis-kin}}

The optical spectroscopy reported by de Vries et al.\
(\cite{devries00}) gives $z=0.3547$ consistently for both
\oiii\,$\lambda4959$ and \oiii\,$\lambda5007$, while
H$\beta\,\lambda4861$ is at $z=0.3536$. Thus, the observed
centroid velocity of the \oiii\ doublet is formally about 325~\kms\
larger than that of H$\beta$. We believe that this difference is
probably significant. The spectral resolution was 15~\AA, but the
centroid wavelengths are listed in integer \AA, i.e with 4 significant
figures. The \oiii\ doublet lines are narrow and clearly have an
excellent signal-to-noise ratio, so the total uncertainty in the
centroid of these lines could well be of order $\pm$1\AA\ in
wavelength, or $\pm0.0002$ in redshift, or $\pm50$~\kms\ in
velocity. The observed FWHM of H$\beta$ looks like it is roughly
50~\AA, or 2500~\kms, and the line has a lower signal-to-noise ratio,
so the uncertainty in its centroid is probably several times larger
than for the \oiii\ doublet. Nevertheless, we think that the combined
velocity uncertainty probably cannot account for the entire 325~\kms\
observed centroid velocity offset between \oiii\ and H$\beta$.

The \oiii\ line velocity and the velocity of the deepest \hi\ features
agree to within 10~\kms\ (see Fig.~\ref{fig:overview}).  The radio
spectral resolution is 4.5~\kms, and the \hi\ absorption features are
present at high signal-to-noise, so the uncertainty in this match is
dominated by the optical data, and we believe that the offset of the
\oiii\ velocity from that of the deep \hi\ feature is not significant.

The VLBI data show that there is atomic gas at the \oiii\ velocity
spanning a projected distance of several hundred parsecs, and probably
overlapping in projection with the radio core. This makes it likely
that the \oiii\ emission is from a conventional narrow line region
(NLR) surrounding the active nucleus, rather than from an isolated
cloud moving with a fairly high peculiar velocity. In
Sec.~\ref{subsec:dis-loc} some less conventional possibilities are
discussed, but we believe these are less plausible. A substantial
centroid offset from the systemic velocity of the host galaxy seems in
general more likely for the broad line region (BLR), which shows the
kinematics of gas under the immediate local influence of the active
galactic nucleus, than for a region with an extent of several hundred
parsecs. Thus, we have adopted the \oiii\ redshift, $z=0.3547$, as the
redshift of the host galaxy, and the zero point of the velocity scale
used for our main analysis. This means that both the H$\beta$ line and
the lower opacity \hi\ absorption features extend to negative
velocities (i.e.\ outflowing, approaching the observer); in the
restframe, the velocity offset of H$\beta$ is $-244$~\kms.

The \hi\ absorbers in \object{2050+364} thus appear to be somewhat different
than in PKS\,1549$-$79 (Tadhunter et al.\ \cite{tadhunter01}) and
B1221$-$42 (Morganti, priv.comm.). In those objects the velocity of the
\hi\ line does not coincide with the optical redshift of the \oiii\ 
line, but rather with the \oii\ line. Because of its lower ionisation
level, one might expect \oii\ to originate in a larger region and
perhaps to be more representative of the systemic velocity than \oiii.

\subsection{The locations of the \hi\ absorbers relative to the radio
source \label{subsec:dis-loc}}

We think that all of the \hi\ absorption may well occur in the same
region as the radio source, in the inner kiloparsec of the host galaxy,
and that the presence of this atomic gas might have some connection
with the active nucleus, or with the radio morphology. Below we argue
that an association of the narrow, deep \hi\ component with the NLR in
\object{2050+364} is plausible, and that the broader \hi\ absorption towards
region W could be related to the BLR\null.

The atomic gas is probably not kinematically disturbed very much as a
result of interaction and bending of the jet at E\null. The absorption
profile there only has two narrow components, and moreover, the
dominant deep line is rather similar to the one at W, where no bending
of the jet is visible. If the jet at E does impact and disturb clouds
containing atomic hydrogen, these probably do not cover the radio
source as seen along our line-of-sight.

The sharp, deep absorption features at low velocity are at first sight
the most puzzling to explain. The integrated peak opacity in \object{2050+364},
16~\%, is the third deepest in the sample of 41 CSS and GPS sources
analysed by Pihlstr\"om et al.\ (\cite{pihl03}). Based on the
anti-correlation of peak opacity with linear size in that sample,
$\le1$~\%\ absorption would be more typical for a projected extent of
$\sim$300~pc. Conversely, the FWHM of the integrated deep \hi\
absorption feature in \object{2050+364}, 16~\kms, is the narrowest in the
sample, where most of the FWHM are in the range of fifty to several
hundred \kms, with no obvious dependence on source linear size.

Because the remarkably high peak opacity is partly balanced by a
remarkably low FWHM, the integrated column depth lies only mildly above
the anti-correlation with linear size in the sample of Pihlstr\"om et
al.\ (\cite{pihl03}): assuming uniform coverage and a spin temperature
$T_{\mathrm sp}\sim100$ K, as is often thought to prevail in typical
ISM conditions, we obtain $N({\mathrm
\hi})\sim5\times10^{20}(T_{\mathrm sp}/100)$~cm$^{-2}$; column depths
for the individual features are listed in Table~\ref{table:linegauss}.

The most likely location for this \hi\ absorption is in the neutral
cores of the clouds in the NLR, given the excellent correspondence with
the velocity centroid of the \oiii\ line, and the projected extent of
at least 300~pc.

Taking $n_{\mathrm \hi}=100$~cm$^{-3}$ as a rough estimate of the
atomic gas density in NLR clouds, and using the 300~pc transverse
extent covered by the absorbers as an estimate of the line-of-sight
depth through the NLR as well, this would imply that clouds are present
along about 1~\%\ of the pathlength. The key distinguishing property,
then, is the low kinematic dispersion. The main direction of motion
and/or rotation in the NLR of \object{2050+364} is evidently
perpendicular to our line-of-sight, and coherent in velocity to
$\sim$10~\kms\ over several hundred parsecs both along the
line-of-sight and transverse to it. It is this velocity coherence which
allows a deep, narrow absorption line to build up. Thus, it seems
unlikely that the absorption is related to directly inflowing gas,
feeding the nucleus, because the velocity is so similar over a region
substantially larger than typical accretion regions.

On the other hand, the absorption features of comparatively low opacity
but spanning a broader observed velocity range (to nearly
$-200$~\kms) occur only towards W, which we believe is most likely to
be at or close to the nucleus (see Sec.~\ref{subsec:dis-loc}). It is
plausible that this shows neutral hydrogen which is either being
entrained by the inner jet of \object{2050+364}, or is flowing out from
the accretion region. Perhaps this atomic gas is related to the BLR in
\object{2050+364}, particularly since the H$\beta$ centroid, while
subject to considerable uncertainty, is probably also at a negative
velocity of a few hundred \kms\null. The neutral gas, especially
that part of it which happens to be visible in absorption, is likely to
sample only a part of the velocity profile of the ionised gas. The
broader part of the \hi\ profile clearly has several distinct kinematic
components (a possible representation with Gaussian lines is given in
Fig.~\ref{fig:linegauss} and Table~\ref{table:linegauss}); the average
opacity is $\tau\sim0.025$. Assuming W to be uniformly covered gives a
column depth integrated over the observed velocity range
$-50$~\kms\ to $-180$~\kms, of roughly $N({\mathrm
\hi})\sim6\times10^{20}(T_{\mathrm sp}/100)$~cm$^{-2}$.  Making the
further and surely oversimplified assumption that the absorbing atomic
gas is also uniformly dense along a pathlength comparable to the total
transverse extent of W ($\sim$10~pc), then the atomic gas density would
be $n_{\mathrm \hi}=20$~cm$^{-3}$, which is rather low. Unfortunately,
with the EVN we cannot resolve region W at the frequency of the \hi\
line. The optical BLR itself probably has an extent of at most 1~pc, so
if the atomic gas were also confined to that region only, then its
density would be perhaps two or three orders of magnitude higher,
because the pathlength would decrease, the opacity and column depth
would increase since the covering factor of W would be less than unity,
and finally, close to the active nucleus the spin temperature could be
significantly elevated.

Could the \hi\ absorption arise instead outside the host galaxy of
\object{2050+364} ? A chance superposition, in which the galaxy visible at the
position of \object{2050+364} would be a foreground object rather than the host
of the radio source, is {\it a priori} unlikely. Such a situation does
exist for lensed radio sources, such as 0218+357, where the foreground
lensing galaxy also leads to \hi\ absorption (Carilli, Rupen, \&\ Yanni
\cite{carilli93}), but we do not think that the radio morphology of
\object{2050+364} is suggestive of gravitational lensing.

Close to the radio position, two other galaxies of comparable
brightness are visible in the images of Biretta et al.\
(\cite{biretta85} and O'Dea et al.\ (\cite{odea90}). One or both of
these could be a true companion to the host galaxy of the radio
source. The closer neighbour, in PA 180\dgr, is centred at a projected
distance of only 2~arcsec (10~kpc at $z=0.3547$), and so part of its
disk could well be overlapping with the radio source in projection. The
other one, centred at a projected distance of 7~arcsec (35~kpc at
$z=0.3547$) in PA $-120$\dgr, seems to show an extension (a spiral arm
?) in the direction of the radio source, and this might be another
candidate foreground absorber.

However, in any external absorber scenario considered, that other
galaxy would probably have to be responsible for both the \oiii\ line
as well as the \hi\ absorption, given the velocity correspondence to
within 10~\kms\null. But that would mean that the other galaxy may also
have an active nucleus, or prodigious star-bursting activity, since
strong \oiii\ emission is not usually seen in ordinary galaxies. The
optical spectrum shown in de Vries et al.\ (\cite{devries00}) suggests
a high ratio of \oiii\ to H$\beta$, i.e., high ionisation gas. This
would indicate ionisation by an AGN rather than a starburst. In our
view the required coincidence of two active galaxies makes external
absorber options much less plausible than the simple model discussed
earlier that the \hi\ absorption shows atomic hydrogen in the NLR and
BLR associated with the compact radio source \object{2050+364}.

\section{Summary \label{sec:sum}}

We have presented and discussed VLBI continuum and spectral line data
at 1049~MHz of the compact radio source \object{2050+364}, and we have
interpreted the multi-frequency continuum VLBI images of LF. Our
conclusions can be summarised as follows:

-- The continuum structure at 1049~MHz consists of a faint component N
  to the northeast, in addition to the two well-known components E and
  W, which constitute the original double; their flux density ratio is
  3:1.

-- The compact-double properties hold only over a limited
  frequency range. W, E, and N, in that order, have decreasing
  compactness, steepening low frequency spectral index, and decreasing
  peak frequency.

-- Compactness and radio spectra, plus the alignment of
  substructures within W and E, suggest that \object{2050+364} is a one-sided
  core-jet originating at or near W, bending sharply (in projection) at
  E, and continuing towards N.

-- At the high velocity end of the absorption spectrum is a deep \hi\
  line, which reaches a maximum opacity {$\tau=0.26$ at E, with a FWHM
  of 13~\kms. At W it is even narrower, with a FWHM of 10~\kms\ and
  somewhat lower opacity, $\tau=0.21$. At N the opacity is
  $\tau\sim0.09$. This absorption thus covers the entire source,
  extending over $>300$~pc, albeit with significant differences in
  opacity. Implied column depths are $N({\mathrm
  \hi})\sim5\times10^{20}(T_{\mathrm sp}/100)$~cm$^{-2}$}.

-- Extending to lower velocities is lower opacity \hi\ absorption,
  which only covers W\null. The average opacity between $-50$~\kms\ and
  $-180$~\kms\ is $\tau\sim0.025$, and the estimated column depth is
  $N({\mathrm \hi})\sim6\times10^{20}(T_{\mathrm sp}/100)$~cm$^{-2}$.

-- The centroid of the \oiii\ optical doublet lines, $z=0.3547$,
  coincides to within 10~\kms\ with the distinct deep line at the high
  velocity end of the \hi\ radio absorption line spectrum. The formal
  line centroid of the optical H$\beta$ line is at an observed
  velocity of $-325$~\kms\ relative to \oiii.

-- We believe that the uncommonly deep but also uncommonly narrow \hi\
  absorption line is likely to be due to atomic gas in the cores of NLR
  clouds in the inner kpc of \object{2050+364}. Assuming NLR clouds of density
  $n_{\mathrm \hi}=100$~cm$^{-3}$ in a region with a 300~pc radius would
  imply their presence along 1~\%\ of the line-of-sight. The
  direction of motion and/or rotation in the NLR of \object{2050+364} is
  coherent to $\sim$10~\kms\ over several hundred parsecs
  both along the line-of-sight and transverse to it, and is largely
  perpendicular to our line-of-sight.

-- We believe that the lower opacity absorption at W ranging to
  $-200$~\kms\ is likely to be due to atomic gas which is either being
  entrained in the inner parsecs of the jet of \object{2050+364}, or is flowing
  out from the accretion region. It is plausibly related to the BLR in
  \object{2050+364}. The atomic gas density is at least $n_{\mathrm
  \hi}=20$~cm$^{-3}$, for a 10~pc absorbing region, but could well be
  as much as three orders of magnitude higher if the region is smaller.

\begin{acknowledgements}
  The European VLBI Network is a joint facility of European, Chinese,
  South African and other astronomy institutes funded by their national
  research councils. The WSRT is operated by ASTRON (The Netherlands
  Foundation for Research in Astronomy) with support from the
  Netherlands Foundation for Scientific Research (NWO). The data were
  correlated at the NRAO, Socorro processor. This research has made use
  of the NASA/IPAC Extragalactic Database (NED), which is operated by
  the Jet Propulsion Laboratory, California Institute of Technology,
  under contract with the National Aeronautics and Space Administration
  in the United States of America. We thank Dr.\ Raffaella Morganti for
  useful discussions, and the referee, Dr.\ Seiji Kameno, for a
  very thorough report, which has led to considerable enhancements to
  this paper.

\end{acknowledgements}

\end{document}